\documentstyle[editedbook,epsfig,psfig,epsf]{mq}

\def\cm2{cm$^2$ }
\def\se1{s$^{-1}$ }

\begin{opening}
\title{Do High Frequency QPOs Depend on Phase of Low Frequency QPOs in XTE
J1550-564 ?}
\author{W. Yu, M. van der Klis \& R. Fender}
\institute{Astronomical Institute `Anton Pannekoek', University of Amsterdam, 
Kruislaan 403, 1098 SJ Amsterdam, the Netherlands}
\end{opening}

\runningtitle{Workshop Proceedings}
\runningauthor{Yu, van der Klis\& Fender}

\begin{document}
\vspace{-0.5cm}
\begin{abstract}
{\small We have studied the dependence of the high frequency 
QPO ($\sim$ 284 Hz) on the phase of the low frequency QPO 
($\sim$ 6 Hz) in the black hole X-ray binary 
XTE J1550-564 in the observations of the {\it Rossi} X-ray 
Timing Explorer (RXTE) performed on MJD 51241. 
By selecting the local maxima and the local minima in the 
light curve on the 6 Hz QPO time scale, we have found the 
corresponding high frequency QPO frequencies are consistent 
within 1.5 $\sigma$. However, the average central QPO frequency 
of the maxima and the minima is about 2.0 $\sigma$ lower than 
the average high frequency QPO frequency obtained in the entire 
observation. This marginally suggests that the high frequency 
QPOs probably varies in frequency on short time scales. 
We briefly discuss these results and their consequences.}
\end{abstract}

\section{Introduction}
XTE J1550-564 was identified as a Galactic "microquasar" from 
both X-ray and radio observations. It outbursts have been 
observed by the {\it Rossi} X-ray Timing Explorer (RXTE) 
for a few times. High frequency QPO with a frequency as high 
as 285 Hz was observed casually from this source in the outbursts in 
both 1998-1999 (Remillard et al. 1999; 
Homan et al. 2001) and 2001 (Miller et al. 
2001). X-ray state transitions during the rise and the decay 
of its outbursts were also observed (see e.g. Homan et al. 2001). 
On the other hand, superluminal jet production has been observed 
in radio (Hannikainen et al. 2001). 

The high frequency QPOs occur ocassionally in the frequency 
range between 102 Hz and 285 Hz in the 1998-1999 outburst 
(Remillard et al. 2001; Homan et al. 2001) and the 2000 
outburst ( Miller et al. 2001). Miller et al. 2001  found that 
these high frequency QPOs are in fact two QPOs present 
simultaneuously, and with a rough frequency ratio of 2:3. Possible 
evidence for the harmonic origin of the high frequency QPOs has been 
found also in other microquasars such as GRO J1566-40 
(Remillard et al. 2002; see also papers in this proceedings). 
If these QPOs are indeed harmonic-related features, the QPO width 
in frequency can put strong constrains 
on the range that the high frequency QPO possiblely moves. If the QPO 
frequency is related to the general relativistic frequencies at the 
inner accretion disk, then the range of the QPO frequency may 
put constrains on the movement of the inner disk radius. 

\section{Observations, Data Analysis and Results}
RXTE observations of XTE J1550-564 on 4 March 1999 show a strong 
high frequency QPO (HFQPO) at 284 Hz and a strong low frequency QPO 
(LFQPO) at $\sim$ 6 Hz with its harmonics at $\sim$ 12 Hz (see Fig.12 in 
Homan et al. 2001). The HFQPO is always stronger in the hard 
energy band above 6.5 keV (Homan et al. 2001).  In the following 
analysis, we use the entire PCA energy band ($\sim$ 2--60 keV) to 
study the LFQPO and the energy band above 6.5 keV (6.5--60 keV) to 
study the HFQPO.

We first select time intervals corresponding to the local maxima and 
the local minima in the 2-60 keV light curve with a window of 1.25 
seconds and a time resolution of 0.03125 seconds. Then we require the 
local maxima and the local minima to be related to the 
variation of 2--12 Hz. A total of $\sim$ 2400 pairs of local maxima and 
minima were obtained, corresponding to a coverage of 1/9 of the entire 
data set ($\sim$ 3600 s). 

We then calculate the Fourier power spectra corresponding the local 
maximal and local minimal intervals and the overall observations 
in the 6.5--60 keV band. The HFQPOs were studied and compared.  
The average HFQPO frequency corresponding to the minima is 283.8$\pm$1.7 Hz, 
while the average HFQPO frequency corresponding to the maxima is 
277.6$\pm$3.6 Hz. In the entire observations, the average HFQPO is at 
284.9$\pm$0.9 Hz. The frequency difference between that in the maximal 
and that in the minimal intervals is about 1.5 $\sigma$, and the 
frequency difference between that in the extream intervals 
(280.7$\pm$2.0 Hz) and that 
in the entire observation is about 2 $\sigma$. The average power 
spectrum corresponding to the maxima is shown in Fig.1. The 
278$\pm$4 Hz QPO, together with its 2:3 harmonic-related QPO at 
184$\pm$4 Hz, and a feature which is not significant at 92$\pm$2 Hz 
are shown. 
\begin{figure}[htb]
\centering
\psfig{file=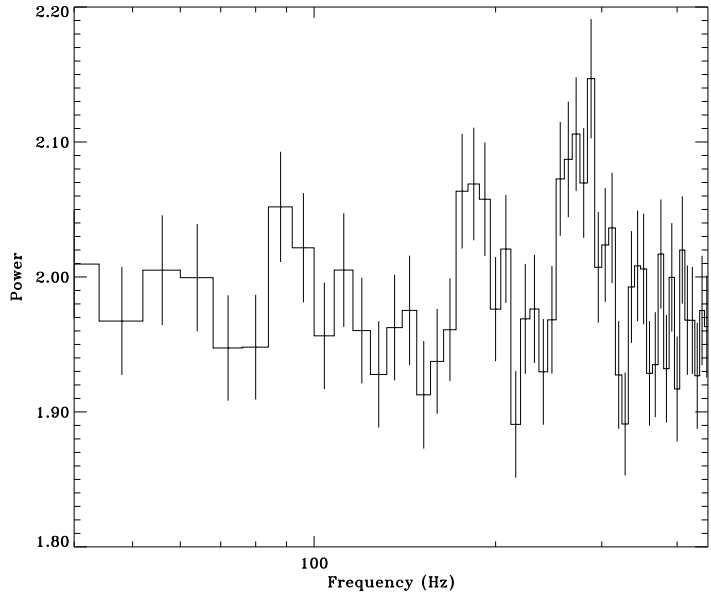,width=10cm}
\caption{The average power spectrum corresponding to the maxima selected 
for the $\sim$ 6 Hz QPO pulses.  }
\label{fig:1}
\end{figure}

\section{Are High Frequency QPOs Constant ?}
From the above analysis, we show that within time scales as short as that 
corresponding to the LFQPOs (e.g $\sim$ 0.125s), the HFQPO frequency 
varies at a significance level of 2 $\sigma$. This suggests that the HFQPOs 
are probably not constant in frequency on short time scales. The QPO frequency 
movement on short time scales can contribute to part of the HFQPO width 
observed in each individual observations. 
 
The movement of the HFQPO frequency in the microquasars is also suggested 
by the HFQPO frequency distribution in XTE J1550-564 and GRO J1655-40 
(Remillard et al. 2002). In the histogram of HFQPO frequencies for 
XTE J1550-564 and GRO J1655-40, the entire width of the HFQPOs or their 
harmonics is about 50--80 Hz, much larger than the HFQPO width in 
individual RXTE observations. This suggests that the HFQPO frequency 
moves also on longer time scales.  

\section{Discussion}
The frequencies of the HFQPO in black hole X-ray binaries are thought 
to be related to some fundamental frequencies at the innermost 
orbit around the black hole, e.g. combinations of the azimuthal and 
radial coordinate frequencies in general relativity (Strohmayer 2001), 
diskoseismic oscillation frequency (Wagoner 1999) or resonance frequency 
between the Keplerian and radial coordinate frequencies (bramowicz \& 
Kluzniak 2001).  If this is true, then the movement of the HFQPO 
frequencies probably suggests that the radius of the innermost orbit 
vary. An important consequence is that the maximal fundamental 
HFQPO frequency is the central HFQPO frequency plus its half width, about 
25--40 Hz higher than the HFQPO central frequency which was used 
in constraining the mass and the spin of the central black hole.

\section*{Acknowledgments}
This work was supported in part by the Netherlands Organization 
for Scientific Research (NWO) under grant 614.051.002. W. Y. is 
on leave from the Institute of High Energy Physics (IHEP) of 
Chinese Academy of Sciences and would like to acknowledge partial 
support from NSFC. This work has made use of data obtained through 
the High Energy Astrophysics Science Archive Research Center Online 
Service, provided by the NASA/Goddard Space Flight Center.

\end{document}